\documentclass[conference]{IEEEtran}
\usepackage{floatrow}
\RequirePackage{inputenc}          

\RequirePackage[T1]{fontenc}       
\RequirePackage[english]{babel}   

\RequirePackage{amsmath}    
\RequirePackage{amssymb}    
\usepackage{amsthm}

\RequirePackage{array}
\usepackage{multirow}

\RequirePackage{hhline}

\usepackage{graphicx,dblfloatfix}
\usepackage{epsfig}

\newtheorem{example}{Example}

\newtheorem{lemma}{Lemma}

\newtheorem{theorem}{Theorem}
\newtheorem{notation}{Notation}
\newtheorem{remark}{Remark}

\newtheorem{conjecture}{Conjecture}

\usepackage{bbm}

\long\def\symbolfootnote[#1]#2{\begingroup%
\def\thefootnote{\fnsymbol{footnote}}\footnote[#1]{#2}\endgroup}

\def\nmin{n_{\min}}
\def\nmax{n_{\max}}
\def\Lmax{L_{\max}}
\def\NM{{\cal N}_{M}}
\def\NMvec{{\cal N}_{\mathbf M}}

\newcommand{\binomial}[2]{\left(\begin{array}{c} #1\\#2 \end{array}\right)}

\def\prob{{\color{black}{\mathbb{P}}}}

\begin{document}
\title{On Distributed Storage Allocations for Memory-Limited Systems} 

\author{\IEEEauthorblockN{Iryna Andriyanova}
\IEEEauthorblockA{ETIS group, UMR-8051\\ ENSEA/University of Cergy-Pontoise/CNRS\\ Cergy-Pontoise, France\\
Email: iryna.andriyanova@ensea.fr}
\and
\IEEEauthorblockN{Pablo M. Olmos}
\IEEEauthorblockA{Departamento Teor\'ia de la Se\~nal y Comunicaciones\\
Universidad Carlos III de Madrid\\
Madrid, Spain\\
Email: olmos@tsc.uc3m.es}
}

\maketitle

\begin{abstract}
In this paper we consider distributed allocation problems with memory constraint limits. Firstly,  we propose a tractable relaxation to the problem of optimal symmetric allocations from \cite{distr-alloc:it2012}. The approximated problem is based on the Q-error function, and its solution approaches the solution of the initial problem, as the number of storage nodes in the network grows.  Secondly, exploiting this relaxation, we are able to formulate and to solve the problem for storage allocations for memory-limited DSS storing and arbitrary memory profiles. Finally, we discuss the extension to the case of multiple data objects, stored in the DSS.
\end{abstract}


\section{Introduction}

In last years, more and more attention is given to wireless distributed storage systems, or the so called wireless caching networks, assumed to deal with the problem of the network bandwidth bottleneck in future-generation wireless networks, due to the increase of the wireless data traffic related to such applications as on-line video streaming, web browsing etc.   
It is worth mentioning that the nowadays wireless networks have more and more of available network bandwidth, thanks to new communication technologies, and also to the fact that the cell size continues to decrease. This implies that one of the next problems to be considered in the distributed storage context is more related to the {\it limitation on the amount of storage memory}, available in the system, rather than to network parameters of the system. For instance, the memory limitation can appear in following situations:
1) when the amount of data to store is very important (i.e. in order to improve the service of on-line video streaming, a large choice of video files is proposed to a user); 2) when the data, related to some application, is stored over the user devices (i.e. in a Device-to-Device communication network), while the device memory, reserved by this application, is limited; 3) in the multi-user scenario with a large number, the the data is stored with a high redundancy, thus improving the quality of experience (QoE) perceived by the users, but also leading to large memory volumes stored in the network. 

Therefore, in this work we focus on memory-limited distributed storage systems. 
We study the problem of storing data objects (files) in a set of storage nodes, each of them having some \textit{maximum memory volume}, available for use. 
For simplicity, it is assumed that all storage nodes can be accessed successfully with the same probability $p$.  
We consider the problem of maximizing the probability of success recovery, and we aim to characterize the optimal storage allocation, given the memory profile of the system and both number and sizes of stored files. 

The problem is a generalization of a storage allocation problem with unlimited memory, considered in \cite{distr-alloc:it2012}.
However, if one directly extends the optimization problem of \cite{distr-alloc:it2012} to the memory-limited case, it becomes difficult to handle.
The reason for that is that the main objective function, on which the result from \cite{distr-alloc:it2012} is based,  is in fact a complimentary cdf of a binomial distribution ${\cal B}(n,p)$ with parameters $n$ and $p$.
This objective function is discrete and non-monotone, and its analysis is already tedious in the original setting of \cite{distr-alloc:it2012}. 
So, it needs to be handled very carefully in the memory-unlimited case, which is even more involved. 
Therefore, before addressing the case with limited memory, we make our first contribution by defining a relaxation of the initial optimization problem from \cite{distr-alloc:it2012} by using a continuous approximation. \color{black}{} We use the fact that the probability function $f_{{\cal B}(n,p)}(i)$, $i\in\mathbb{N}_{0}$, of the binomial distribution ${\cal B}(n,p)$ can be written as
\begin{align}\label{approx}
f_{{\cal B}(n,p)}(i)=\frac{1}{\sqrt{2\pi n p (1-p)}}\text{e}^{-\frac{(i-np)^2}{\sqrt{2\pi n p (1-p)}}}\left[1+\mathcal{O}(\frac{1}{\sqrt{n}})\right],
\end{align}
which is the normal distribution probability function with parameters $\mu = np$ and $\sigma = \sqrt{np(1-p)}$, up to corrections that vanish as $n\rightarrow\infty$.
\color{black}
%
Based on this result, we propose to relax the objective function using a Q-error function 
$$Q(x) = \frac{1}{\sqrt{2 \pi}}\int_x^{\infty} e^{-t^2/2}d t, \quad x\in {\mathbb R^+}.$$ 

The approximation is accurate even for moderate values of $n$ (of order several dozens of nodes present in the storage network, see Section \ref{sec:memory-unlimited}). 

Thanks to the proposed relaxation, we can treat the memory-limited storage case. Our second contribution is in defining a tractable optimization problem 
in the case of an arbitrary memory profile of the network and in characterizing the optimal storage allocation in this case.
In particular, we define a relaxed optimization problem in the memory-limited case and solve it for the case when there is only one data file stored in the network (see Section \ref{sec:memory-limited}). 
Moreover, a conjecture on the case of two stored data objects is developed in Section \ref{discussion}, thus opening the problem of storing multiple data objects in a memory-limited distributed storage system. This our third and last contribution.


\section{Optimal Symmetric Allocations Revisited}
\label{sec:memory-unlimited}
\subsection{System Model and State of the Art}

Assume a distributed storage system with $N$ storage nodes.
A source stores a data object of normalized unit size that is encoded and stored in a distributed manner over the system, subject to a given total storage budget $T$ ($T$ is the inverse of the rate of the underlying code). 
Let $x_i$ be the amount of coded data stored in node i, $1\le i\le N$. Then, 
\begin{align}
\label{eq:xi}
\sum_{i=1}^N x_i \le T.
\end{align}

The data collector wishes to download and to recover (i.e. to decode) the stored data object. It is assumed that it accesses each of $N$ storage nodes independently with some access probability $p$.
One aims therefore to find an optimal allocation $X=(x_1, x_2,\ldots,x_N)$, subject to (\ref{eq:xi}), so that the probability to recover the data successfully is maximized. Assuming that the data was encoded using a MDS code, the data object can be recovered if the amount of data collected by the collector is above or equal to one unit. 
This is translated into the following optimization problem:

\vspace{-0.2cm}
\begin{align}
X^*= \sup_{X} \sum_{{\cal S}\in \mathcal{P}(\{1,\ldots,N\})} p^{|{\cal S}|} (1-p)^{N-|{\cal S}|} \color{black}\mathbbm{1}\left( \sum_{i\in {\cal S}} x_i \ge 1\right), 
\label{eq:P0}
\end{align} 
subject to (\ref{eq:xi}), where  $\mathcal{P}(\{1,\ldots,N\})$ is the power set of $\{1,\ldots,N\}$ and $\mathbbm{1}[\cdot]$ is the indicator function.
This optimization problem can be simplified if the search is restricted to the set of symmetric allocations, i.e. if $X^*$ is assumed to belong to the following subset ${\cal X}$:
\begin{align}
{\cal X} = \cup_{n=1}^N {\cal X}_n \text{ with }
{\cal X}_n =\{X :  \sum_{i=1}^N x_i = T \text{ and }  x_i \in \{0, \frac{T}{n}\} 
\}.
\end{align}

Given that the collector accesses nodes uniformly at random, the probability of recovery (objective function) will not depend on the exact indexes of non-zero allocations but rather  on the number of nodes $n$ used to symmetrically store the data object. \color{black}As each of the $n$ nodes used for storage is accessed with probability $p$,  the probability of successful recovery for a given $n$ value is given by \cite{distr-alloc:it2012}
\begin{align}
\label{eq:B}
\sum_{i= \left\lceil\frac{n}{T} \right\rceil }^n f_{{\cal B}(n,p)}(i)= \sum_{i= \left\lceil\frac{n}{T} \right\rceil }^n {\binomial n i} p^i (1-p)^{n-i}.
\end{align}
and the optimization problem reads as follows:
\begin{align}
(P1): \quad  & n^*= \sup_{1 \le n \le N} \left(\sum_{i= \left\lceil\frac{n}{T} \right\rceil }^n f_{{\cal B}(n,p)}(i) \right), 
\label{eq:P1}
\end{align} 
where the set of corresponding optimal symmetric allocations is ${\cal X}_{n^*}$. \color{black}
The expression above represents a discrete non-monotone function in $n$. In order to find its supremum, it is sufficient to restrict to the following subset of values of $n$:
\begin{align}
\label{eq:setN}
{\cal N} =\{\lfloor T \rfloor, \lfloor 2T \rfloor, \ldots, \lfloor LT \rfloor, N\},
\end{align}
where $L=\left\lfloor \frac{N}{T} \right\rfloor$.
This fact is also stated in equation (9) of \cite{distr-alloc:it2012}.
Using the fact, the authors of \cite{distr-alloc:it2012} obtain the following result (Theorems 3 and 4 in \cite{distr-alloc:it2012}):
\begin{align}
\label{eq:result-dimakis}
n^* = \begin{cases}
\lfloor T \rfloor, & \text{ if } p\lfloor T\rfloor<1;\\
\lfloor LT \rfloor \text{ or }N, & \text{ if } (1-p)^{\lfloor T \rfloor}+2 \lfloor T \rfloor p(1-p)^{\lfloor T \rfloor-1}\le 1.
\end{cases}
\end{align}

The interesting regime of parameters for practical applications is $pT \ge 1$ as in this case the success probability is unbounded from $1$. In the section below, we propose a more tractable optimization problem based on a relaxation to (P1) which  gives a good estimate of solutions regions for $n^*$.

\subsection{ A Q-Function Approximation Applied to (P1)}

Let $\mu = np$ and $\sigma = \sqrt{np(1-p)}$, for some\footnote{Note that here $n$ is a real value and not an integer as before. However, as it corresponds to $n$ above, we keep call it $n$. To avoid the abuse of notation later on, it will always be mentioned if $n \in {\mathbb R}$.} $n \in {\mathbb R}$ such that $n \in [0,N]$ and $p\in[0,1]$. 
Define the optimization problem (P2) as:

\vspace{-0.5cm}
\begin{align}
(P2) \ : \quad & n^* =\sup_{n \in {\mathbb R}} Q\left( \frac{ \left\lceil \frac{n}{T} \right\rceil -\mu}{\sigma} \right), \qquad 1 \le n \le N
\label{P2} 
\end{align} 

\begin{figure}[b!]
\includegraphics[scale=0.52]{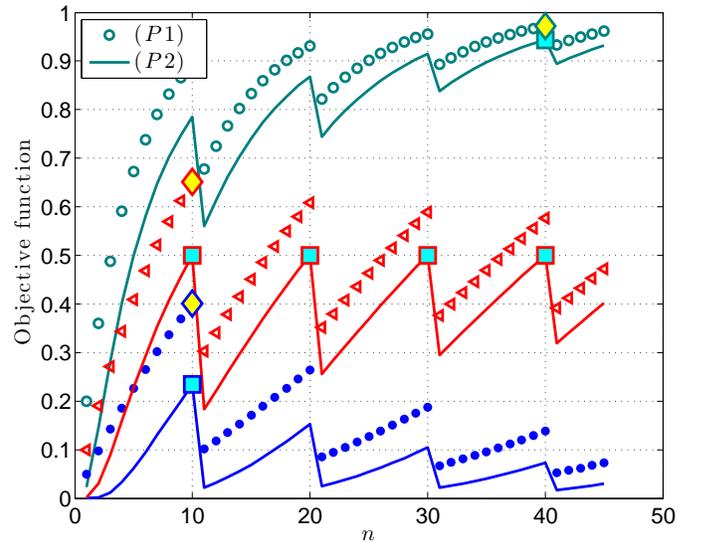}
\caption{\label{Qplots} Objective function (probability of successful recovery) of problems $(P1)$ and $(P2)$ as a function of $n$ for $N=45$ and various cases of $pT$. From top to bottom: $p=0.2$ and $T=10$ $(o)$, $p=0.1$ and $T=10$ ($\square$) and $p=0.05$ and $T=10$ $(\bullet)$. For each case, with $(\Diamond)$ markers we show the solution to $(P1)$  with $(\square)$ makers the solution to $(P2)$.  For $pT>1$ and $pT<1$, the solution to both problems is the same, this is not the case for $pT=1$. Note that $(P2)$ for the $pT=1$ case has multiple solutions.  } 
\end{figure}

(P2) is obtained from (P1) by applying the normal approximation to $f_{{\cal B}(n,p)}$ described in \eqref{approx}.
Thanks to well-known properties of the $Q(x)$ function, we can  characterize the set of solutions as follows:

{\theorem \label{theorem:Qapprox}
The solution of (P2) is 
\begin{align}
\label{eq:nstar}
n^* = \begin{cases}
\lfloor T \rfloor, & pT<1 \text{ (Case 1)};\\
\in {\cal N}\backslash N, & pT=1 \text{ (Case 2)};\\
\lfloor LT \rfloor, & \frac{1}{T}<p<\frac{L+1}{N} \text{ (Case 3)};\\
N, & \frac{L+1}{N} \le p \le \frac{L+1}{N-\sqrt{LT}}+\frac{1}{N\sqrt{LT}-\sqrt{T}}\\
& \text{(Case 4)};\\
\lfloor LT \rfloor, &  p>\frac{L+1}{N-\sqrt{LT}}+\frac{1}{N\sqrt{LT}-\sqrt{T}} \text{ (Case 5)}.
\end{cases}
\end{align}
}

The outline of proof of Theorem \ref{theorem:Qapprox} is given in Appendix \ref{sec:proof}.  In Fig. \ref{Qplots}, we show the objective function of problems $(P1)$ and $(P2)$ as a function of $n$ for $N=45$ and different $(p,T)$ pairs. Note that, while in the cases $pT>1$ and $pT<1$, the solution to both problems is the same, this is not the case for $pT=1$. Recall here that the range of interest for distributive storage allocation problems is $pT>1$.  In Table \ref{TABLEALPHA} includes a  measure of the disparity between $(P1)$ and $(P2)$ for $N=10$, $N=20$ and $N=45$. For the  grid $p=0:10^{-3}:1$ and $T=0:0.1:N$ and increasing $N$ values
we compute the following quantities:
\begin{itemize}
\item Fraction $\alpha$ of points in the grid for which \newline $|n^*_{(P1)}= n^*_{(P2)}|$;
\item Fraction $\beta$ of points in the subregion of the grid with $pT>1$ for which $|n^*_{(P1)}= n^*_{(P2)}|$.
\end{itemize}

\begin{table}[t!]
\begin{center}
\begin{tabular}{|c|c|c|}
\hline
Size of DSS & $\alpha$ & $\beta$\\
\hline
$N=10$ & 0.8823 & 0.904  \\
$N=20$ & 0.9048 & 0.9208 \\
$N=45$ & 0.9345 &0.9532 \\
\hline 
\end{tabular}\caption{Measuring the disparity between $(P1)$ and $(P2)$}\label{TABLEALPHA}
\end{center}
\end{table}

Observe that both $\alpha$ and $\beta$ improve  for larger $N$ values, which indicates that our approximation gets tight in the limit $N \rightarrow \infty$. \color{black}{}
The equivalence of (P2) and (P1) for large values of $N$, together with the fact  that the optimal symmetric allocation $n^*$ approaches the optimal (asymmetric) one when $N$ goes to infinity \cite{distr-alloc:it2012}, gives us the asymptotic optimal solution of the distributed storage allocation problem.  As it is easier to deal with (P2), we  apply it to our case of interest which is the distributed storage allocation in memory-limited systems.  

\section{One single data object in Memory-Limited DSS}\label{memory1user}
\label{sec:memory-limited}
Let us consider the DSS of our interest: a storage node $i$ is assumed to have an available memory $M_i$, which can be used to store the coded data. 
Let a data object of total budget $T$ be stored in the DSS.
Note that if $T\le \min_i {M_i}$, then the problem is equivalent to the memory-unlimited case. Also, if $T > \sum_i M_i$, then the allocation solution does not exist. 
So let us focus on an interesting region of $T$ which is $\min_i M_i < T \le \sum_i M_i$.
With some abuse of notation, let the set of memory-limited symmetric allocations of size $n$ be defined as:
\begin{align}\label{sym}
{\cal X}_n =\{X : x_i \le M_i,  x_i \in \{0, \frac{T}{n}\}, \  \#\large(x_i = \frac{T}{n}\large)=n\},
\end{align}
where $\#(x_i=a)$ denotes the number of elements in $X$, equal to $a$.

\subsection{Constant Memory Profile}
We start with developing an intermediate result for a DSS with a constant memory limit $M$. 
Note that the solution to this problem will differ from the unconstrained memory scenario summarized by \eqref{theorem:Qapprox} for those cases where we are interested in storing a large amount of memory in an small set of nodes. Because of the memory limit $M$, for $pT<1$, the symmetric minimum spreading solution might not be optimal. 

We define the set of \textit{quasi-symmetric} allocations of size $n$ and of memory volume $M$ as:\color{black}
\begin{align}\label{quasi}
{\cal X}^M_{n} =&\{X :  x_i \le M,\#(x_i = M)=n-1,\ \#(x_i = R)=1 \},
\end{align}
with $R=T-Mn$. Hence, in such allocation we use the complete memory $M$ in $n-1$ nodes and the rest of the data object, i.e., $R<M$ is stored in an additional node. $n$ nodes are used in total. By \eqref{sym} and \eqref{quasi}, ${\cal X}_i$, $i\in\mathbb{N}$, represents a set of symmetric allocations where $i$ nodes are used for storing and ${\cal X}^M_i$ is a set of quasi-symmetric allocations where $i$ nodes are sused.  \color{black}

 Define  $\nmin = \lceil \frac{T}{M} \rceil $, and let $L_0$ be the smallest integer such that\footnote{Note that $\nmin$ is the minimum number of nodes we can use to store the budget $T$. Since $\nmin$ might not be contained in the set $\mathcal{N}$ in \eqref{eq:setN}, is $\lfloor L_0 T \rfloor$  the possible solution to the problem 
that is closest to $\nmin$. We assume that $L_0<L$. Otherwise the optimal $n^*$ is given by Cases 3,4, and 5 of (\ref{eq:nstar}) if $L_0=L$, and $n^*= N$ if $L<L_0 \le N$.} $\nmin \le \lfloor L_0 T \rfloor$. Finally, define  $\NM = \{\lfloor L_0 T \rfloor, \ldots, \lfloor LT \rfloor, N\}$.  We have the following result:
%
\color{black}
{\lemma 
\label{lemma:M}
Assume a limited-memory DSS, for which $M_1=\ldots=M_N=M.$ Let $p_0$ be the unique solution of the equation 
\begin{align}
\label{eq:pstar}
p \sum_{i= \left\lceil\frac{1-R}{M} \right\rceil }^{\nmin-1} f_{{\cal B}(\nmin-1,p)}(i)+(1-p) \sum_{i= \left\lceil\frac{1}{M}  \right\rceil }^{\nmin-1} &  f_{{\cal B}(\nmin-1,p)}(i)\nonumber\\
-  \sum_{i=L_0 }^{\lfloor L_0 T \rfloor} f_{{\cal B}(\lfloor L_0 T \rfloor,p)}(i)=0.
\end{align}
Then the set of optimal storage allocations, maximizing the success recovery in this case, is approximated by 
\begin{align}
\label{eq:nstarM}
X^* \in \begin{cases}
{\cal X}^M_{\nmin}, & pT<1 \text{ and }p\le p_0 \text{ (Case 1a)};\\
{\cal X}_{\lfloor L_0 T \rfloor}, & pT<1 \text{ and }p> p_0 \text{ (Case 1b)};\\
{\cal X}_n, \ n\in \NM \backslash N, & pT=1 \text{ (Case 2)};\\
{\cal X}_{\lfloor LT \rfloor}, & \frac{1}{T}<p<\frac{L+1}{N} \text{ (Case 3)};\\
{\cal X}_N, & \frac{L+1}{N} \le p \le \frac{L+1}{N-\sqrt{LT}}+\frac{1}{N\sqrt{LT}-\sqrt{T}}\\
& \text{(Case 4)};\\
{\cal X}_{\lfloor LT \rfloor}, &  p>\frac{L+1}{N-\sqrt{LT}}+\frac{1}{N\sqrt{LT}-\sqrt{T}} \text{ (Case 5)}.
\end{cases}
\end{align}
Moreover, $X^*$ from (\ref{eq:nstarM}) approaches to the set of optimal allocations, when $N$ goes to infinity.
}
\color{black}

The outline of the proof of lemma is given in Appendix \ref{app:lemmaM}. Note that cases 3, 4, 5 of (\ref{eq:nstarM}) are equivalent to cases 3, 4, 5 of (\ref{eq:nstar}). Case 2 is also similar, with the only exception that one should consider now $\NM$ instead of $\cal N$.
The only difference from (\ref{eq:nstar}) is therefore in the fact that, when $pT<1$, the minimum symmetric allocation is not always optimal anymore -- there exist values of $L_0$ and of $p$ for which the best allocation is the quasi-symmetric one.

{\example
Let $p=0.1$, $T=1.4$, $M=0.5$ and $N\ge 3$. The best allocation in this case is the quasi-symmetric one with 2 $x_i$'s equal to $0.5$ and one $x_i$ equal to $0.4$.
}

 {\remark
 By using a Taylor expansion of a Q-function, one can also get a tight approximation of $p_0$:
 {\footnotesize
 \begin{align}
\label{eq:pstar-approx}
p_0 \approx \left( \lceil 1/M \rceil -L_0\sqrt{\frac{\nmin-1}{\lfloor L_0T \rfloor}} \right) & \left( \nmin-1 - \sqrt{(\nmin-1) \lfloor L_0T \rfloor} \right.
\nonumber \\ \left.+ \lceil 1/M \rceil -\left\lceil \frac{1-R}{M} \right\rceil  \right)^{-1}.&
\end{align}
}
 }

\vspace{-5mm}
\subsection{Arbitrary Memory Profile}

\begin{figure*}[t!]
\includegraphics[scale=0.9]{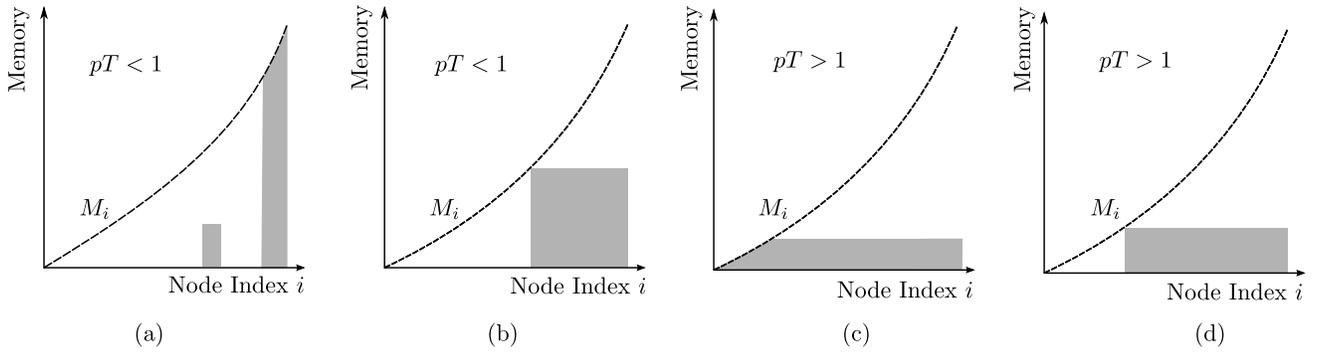}
\caption{Possible storage allocations for an arbitrary memory profile: (a) FLmin allocation; (b) symmetric minimal spreading allocation; (c) ANmax allocation; (d) symmetric maximum-spreading allocation. }\label{arbitrary_allocations}   
\end{figure*}

Now consider an arbitrary memory profile $\mathbf{M}=(M_1, \ldots, M_N)$. 
W.l.o.g., let $M_1 \le M_2 \le \ldots \le M_N$. For this scenario,  two possible  optimal allocations have to be considered for the case $pT<1$ and another another two for the case $pT>1$. We sketch these four scenarios in Fig. \ref{arbitrary_allocations}. When $pT<1$, the  full-load  minimum-support asymmetric allocation, or FLmin allocation for short, uses the complete memory of the nodes with largest available memory. A small fraction of residual data can be stored in any of the remaining nodes in the network. The FLmin allocation is sketched in 
Fig. \ref{arbitrary_allocations}(a). Alternatively, we can store the data object using a symmetric minimal spreading allocation,  see Fig. \ref{arbitrary_allocations}(b). For $pT>1$, the all-node maximum support allocation, or ANmax allocation for short, uses all nodes in the system, defining a quasi-symmetric allocation. The ANmax allocation is is sketched in Fig. \ref{arbitrary_allocations}(c). Also, a symmetric maximum-spreading allocation can be used,  Fig. \ref{arbitrary_allocations}(d).


\subsubsection{Arbitrary Memory Profile with $pT<1$}

{\notation
Let $\nmin$ be the smallest integer such that $\sum_{i=0}^{\nmin-1} M_{N-i}\ge T$.
As before, let $L_0$ be the smallest integer such that $\nmin \le \lfloor L_0 T \rfloor$.
}

We formulate now a necessary condition for FLmin allocation, see Fig. \ref{arbitrary_allocations}(a),  to be a better allocation than the symmetric allocation:

{\lemma \label{lemma:M-any-pT<1}
Denote a FLmin allocation by $X^{FL}_{\nmin}$ of non-zero support $\nmin$ to be
{\footnotesize
$$X^{FL}_{\nmin}=(0, \ldots,0,T-\sum_{i=N-\nmin-2}^{N} M_{i},M_{N-\nmin-2},\ldots,M_{N-1},M_N).$$
}
Then, $X^{FL}_{\nmin}$ is  optimal if
\begin{align}
\label{eq:condition}
m> \frac{T}{\nmin},
\end{align} 
where
\begin{align}\label{mean}
m=\sum_{i=1}^{N} \frac{M^2_i}{\sum_{j=1}^{M} M_j}.
\end{align}
For $m<T/\nmin$, then the symmetric minimal spreading allocation,  Fig. \ref{arbitrary_allocations}(b),  has higher recovery probability. Besides, all modifications to $X^{FL}_{\nmin}$ defined by putting the remainder of the memory in a different set of positions, see Fig. \ref{arbitrary_allocations}(a), achieve the same recovery probability. Denote this set by ${\cal X}^{FL}_{\nmin}$.
}

\color{black}{}

The proof of Lemma \ref{lemma:M-any-pT<1} is given in Appendix \ref{app-any-M-pt<1}. Lemma \ref{lemma:M-any-pT<1} indicates us that depending on the memory profile, one might better to chose either FLmin allocation  or symmetric, minimal-spreading allocation. 

\vspace{1mm}
\subsubsection{Arbitrary Memory Profile with $pT<1$}

{\notation
Let $\nmax$ be the largest integer such that $\frac{T}{\nmax} > M_{N-\nmax}$.
Also, let $\Lmax$ be the largest integer such that\footnote{We assume that $\Lmax > L_0$, 
otherwise $n^*$ will not exist.} $\nmax \ge \lfloor \Lmax T \rfloor$.
}


{\lemma \label{lemma:M-any-pT>1}
Denote the ANmax allocation $X^{AN}_{N}$ to be 
{\footnotesize
$$X^{AN}_{N}=(M_1,\ldots, M_{N-\nmax}, a,\ldots, a),$$
}
where  $a=(T-\sum_{i=1}^{N-\nmax}M_i)/\nmax$. Then $X^{AN}_{N}$ is  optimal if
\begin{align}
\label{eq:condition2}
m (N-\nmax) > \sum_{i=1}^{N-\nmax}M_i,
\end{align} 
where $m$ is defined in \eqref{mean}. If \eqref{eq:condition2} is not verified, then the symmetric maximum-spreading allocation,  Fig. \ref{arbitrary_allocations}(d), has higher recovery probability.
}

%

Proof of Lemma \ref{lemma:M-any-pT>1} is given in Appendix \ref{app:M-any-pT>1}. Lemma \ref{lemma:M-any-pT>1} shows that depending on the memory profile, one might better to chose either AN-max allocation $X^{max}_N$ or symmetric, maximum-spreading allocation. \color{black}{} 

Putting the results of Lemmas \ref{lemma:M-any-pT<1} and \ref{lemma:M-any-pT>1} together, we can state the following:

{\conjecture
Assume a DSS with an arbitrary memory profile with $M_1\le M_2 \le \ldots \le M_N$.
Let $\NMvec = \{ \lfloor L_0 T \rfloor, \ldots, \lfloor \Lmax T \rfloor\} $.
Then

\begin{align*}
X^* \in \begin{cases}
{\cal X}^{FL}_{\nmin}, & pT<1 \text{ and (\ref{eq:condition}) does not hold}; \\
{\cal X}_{\lfloor L_0 T \rfloor}, & pT<1 \text{ and (\ref{eq:condition}) holds};\\
{\cal X}_n, \ n\in \NMvec, & pT=1;\\
{\cal X}_{\lfloor \Lmax T \rfloor}, & pT>1 \text{ and (\ref{eq:condition2}) holds};\\
X^{AN}_N, & pT>1 \text{ and (\ref{eq:condition2}) does not hold}.
\end{cases}
\end{align*}
$X^*$ above approaches to the set of optimal allocations, when $N$ goes to infinity.

}


\section{Extension: When More Than One Data Object Is Stored in the DSS}\label{discussion}

In this work we have analyzed memory-limited DSS systems for a single user. Most of our derivations are based on a tractable approximation proposed for the unlimited memory case. Given the above results, our future interest is to address the multiuser case. We conclude the paper by briefly discussing  the case of two data objects to store ($K=2$). Let the objects have total budgets $T_1$ and $T_2$ respectively.
Also, let the probability that the data collector downloads the file $1$ be denoted by $p_1$, and that he downloads the file $2$ -- by $p_2=1-p_1$. We assume the system is memory-limited. Note that if $\frac{T_1+T_2}{N}\le \min_i {M_i}$, then the problem is equivalent to the memory-unlimited case. Also, if $T_1+T_2 > \sum_i M_i$, then the allocation solution does not exist. 
So, the interesting interval of $T$'s is when $N \min ({\mathbf M}) < T_1+T_2 \le \sum_i M_i$.
 Let the allocations for files $1$ and $2$ be denoted by $X_1$ and $X_2$. If we restrict to symmetric allocations with support $n_1$ and $n_2$ respectively, then  the problem  can be approximated by solving the following optimization problem. 
\begin{align*}
(P4) : &  \sup_{({\cal X},{\cal X})} p_1 Q\left( \frac{ \left\lceil \frac{n_1}{T_1} \right\rceil -\mu_1}{\sigma_1} \right)+p_2 Q\left( \frac{ \left\lceil \frac{n_2}{T_2} \right\rceil -\mu_2}{\sigma_2} \right),
\\
& 
\text{with }\mu_i = n_i p; \ \sigma_i=\sqrt{n_i p (1-p)}, \ \text{ for }i=1,2. \nonumber
\end{align*} 
The following can be proven about the solution of $(P4)$:
\begin{itemize}
\item W.l.o.g., let $p_1>p_2$. To maximize $(P4)$,  we first allocate object 1 given the memory profile $M_1,M_2,\ldots,M_N$ using the results presented in Section \ref{memory1user}. Then, we allocate object 2 using the residual memory profile.
%
%
%
%
\item Let $p_1=p_2=p$. In this case, the game theory suggests that three optimal strategies are possible: a) start allocating the first data object and the allocate the second one; b) to proceed in the inverse order; c) to allocate data of two objects in the mixed way. 
\end{itemize}
\color{black}
%

\section{Conclusion}
In this paper, we have considered a memory-limited DSS, storing one or two data files. 
The memory profile of the system is assumed to be arbitrary; this case is therefore treated in all its generality. 
We wish to emphasize two following points:
\begin{itemize}
\item For a memory-limited DSS, the optimal storage allocation is not necessarily a symmetric one, even in for large network size $N$, even for a constant memory profile when all nodes have the same amount of memory available for storage.
This differs from the result obtained in the usual memory-unlimited case, where the optimal allocation is a symmetric one, in the limit of large $N$.  
\item The obtained result, obtained for the access probability $p$, can be combines with the result from \cite{heterog:isit12}, developed for heterogenous storage networks. Thus it is possible to characterize optimal allocation solutions for memory-limited, heterogeneous DSS.
\item MDS codes, used to prove our results, are the most storage-efficient erasure-correcting codes, but they are not efficient complexity-wise, which makes them impractical to use. It would be interesting to consider a more practical code solution and to check how the optimal allocation changes for this case.
\end{itemize} 

\appendices
\section{Outline of the Proof of Theorem \ref{theorem:Qapprox}}
\label{sec:proof}

First, the following lemma is stated (its proof is quite straightforward and is omitted for the sake of space):
{\lemma
\label{lemma:set-solutions}
Let $L=\left\lfloor \frac{N}{T} \right\rfloor$ as previously.
Then the solution of (\ref{P2}) belongs to the set ${\cal N}$, given by (\ref{eq:setN}).
}
Next, define
$$ c(n) =Q\left( \frac{ \left\lceil \frac{n}{T} \right\rceil -\mu}{\sigma} \right) \text{ with }\mu=np, \ \sigma=\sqrt{np(1-p)} .$$
Owing to Lemma \ref{lemma:set-solutions},  (\ref{P2}) can be written as
$\sup_{n\in {\cal N}}c(n).$ 
To find a solution of this problem, three possible cases are to consider: $pT<1$, $pT=1$ and $pT>1$. 
We discuss only  the latter case, $pT>1$. The first two  cases can be analyzed similarly and their result is stated in stated directly in (\ref{eq:nstar}).

When $pT>1$, $c(\lfloor iT\rfloor )$ is increasing with $i$. and 
$\sup_{i \in \{1, \ldots, L\}} c(\lfloor iT\rfloor ) = L$. Therefore, depending on the value of $N$, $n^*$ is either $\lfloor LT \rfloor$ or $N$.
Note that $N=\lfloor LT \rfloor$ is a trivial case ($n^*=N$).
So let $N>\lfloor LT \rfloor$ and
consider
\begin{align}
\label{eq:difference}
c\left(\lfloor LT\rfloor \right)-c(N) = Q\left(\frac{\sqrt{L }(1- Tp)}{\sqrt{Tp(1-p)}} \right) - Q\left( \frac{ L+1 -Np}{\sqrt{Np(1-p)}} \right)
\end{align}
Two cases are to be distinguished:
\begin{itemize}
\item[a)] $Np\ge L+1$, for which $p \ge \frac{L+1}{N} >\frac{1}{T}$: 
The expression (\ref{eq:difference}) is positive, if the following is satisfied: 
$$- \sqrt{L }\frac{ Tp-1}{\sqrt{Tp(1-p)}} > \frac{ L+1 -Np}{\sqrt{Np(1-p)}}.$$
This holds for 
\begin{align}
\label{eq:condition-p}
p>\frac{L+1}{N-\sqrt{LT}}+\frac{1}{N\sqrt{LT}-\sqrt{T}}.
\end{align}
So, under the condition above, $n^*= \lfloor LT\rfloor $.
Note that, if some $p$ satisfies (\ref{eq:condition-p}), then it also satisfies $p \ge \frac{L+1}{N}$. 
\item[b)] $\frac{1}{T}<p <\frac{L+1}{N}$: it can be verified that, for any value of $p$, $c\left(\lfloor LT\rfloor \right) > c(N)$ and thus $n^*=\lfloor LT\rfloor$.
\end{itemize}

\section{Outline of the Proof of Lemma \ref{lemma:M}}
\label{app:lemmaM}
To show cases 2, 3, 4, 5  the proof is similar to the one of Theorem \ref{theorem:Qapprox}. 
The only difference now is that the solution should belong to $\NM$ instead of ${\cal N}$.
However, for case 1, the symmetric allocations ${\cal X}_{\lfloor L_0 T \rfloor}$ are not necessarily the best choice.
One can show that it only has to be compared with the quasi-symmetric allocations occupying the smallest number of storage nodes, i.e. with the subvector $(M_1, \ldots, M_{\nmax})$.
It is easy to see that the probability of success recovery for the symmetric and quasi-symmetric minimum spreadings, denoted respectively by $P_S$ and $P_{QS}$, are given by
%
\color{black}
\begin{align*}
P_{S} =& \sum_{i=L_0 }^{\lfloor L_0 T \rfloor} f_{{\cal B}(\lfloor L_0 T \rfloor,p)}(i)\\
P_{QS}=&~ p \sum_{i= \left\lceil\frac{1-R}{M} \right\rceil }^{\nmin-1} f_{{\cal B}(\nmin-1,p)}(i)\\
&+(1-p) \sum_{i= \left\lceil\frac{1}{M}  \right\rceil }^{\nmin-1}   f_{{\cal B}(\nmin-1,p)}(i)
\end{align*}

\color{black}

Moreover, both of them are monotonically increasing in $p$
and $P_{S} (1/T) > P_{QS} (1/T)$ while $P_{S} (0) < P_{QS}(0)$. 
So, there exists a unique parameter $p=p_0$ such that $P_{S}(p_0)=P_{QS}(p_0)$, and one can find it by solving (\ref{eq:pstar}).

\section{Proof of Lemma \ref{lemma:M-any-pT<1}}
\label{app-any-M-pt<1}
We are going to use the Markov inequality, which was also used in considering heterogeneous data allocations in \cite{heterog:isit12}. Assuming an arbitrary allocation $X$ with a non-zero support $n$, the optimization problem is approximated as:
\begin{align}
(P3): \quad  & X^*= \sup_{X \text{s.t. (\ref{eq:xi}) holds}} \ \sum_{k=0}^{n} b(n,p) {\prob}(\sum_{i=1}^k Y_i \ge 1), 
\label{eq:P3}
\end{align}
where $Y_i$ are i.i.d random variables and $Y_i \sim p_X(x)$, 
where $P_X(x)$ is the empirical probability distribution, corresponding to the non-zero support of $X$.  
The approximation here comes from the fact that $Y_i$'s are assumed to be i.i.d,
i.e. here the probability distribution, corresponding to the random choice without replacement, is approximated to the probability distribution, corresponding to the random choice with replacement. 
By Markov's inequality,
$${\prob}(\sum_{i=1}^k Y_i \ge 1) \le {\mathbb E}\left(\sum_{i=1}^k Y_i\right) = k m_X,$$
with $m_X$ being the mean of the distribution $P_X(x)$. Therefore, the objective function in (\ref{eq:P3}) can be approximated by
\begin{align}
m_X  \sum_{k=0}^n n b(n,p) = m_X np.
\end{align}
Note that, if $X$ is a symmetric allocation, this quantity equals to $pT$, as $m_X=\frac{T}{n}$. As $pT<1$, then the probability of success recovery is bounded away from 1. 
However, if $X$ is the full-load allocation with the smallest support $\nmin$ and $m_X>\frac{T}{n}$, the objective function will be bounded by a larger quantity than $pT$. 

\section{Proof of Lemma \ref{lemma:M-any-pT>1}}
\label{app:M-any-pT>1}
With the help of the Markov's inequality as for Lemma \ref{lemma:M-any-pT<1}, 
one obtains that the probability of success recovery is upper bounded by 
$p(T-\sum_{i=1}^{N-\nmax} M_i + \tilde m (N-\nmax))$
in the all-node maximum-spreading case,
and by 
$pT$
in the symmetric, maximum spreading case. Hence, the condition (\ref{eq:condition2}) follows.

\bibliographystyle{IEEEtran}
\bibliography{data-allocation}

\begin{thebibliography}{1}
\providecommand{\url}[1]{#1}
\csname url@samestyle\endcsname
\providecommand{\newblock}{\relax}
\providecommand{\bibinfo}[2]{#2}
\providecommand{\BIBentrySTDinterwordspacing}{\spaceskip=0pt\relax}
\providecommand{\BIBentryALTinterwordstretchfactor}{4}
\providecommand{\BIBentryALTinterwordspacing}{\spaceskip=\fontdimen2\font plus
\BIBentryALTinterwordstretchfactor\fontdimen3\font minus
  \fontdimen4\font\relax}
\providecommand{\BIBforeignlanguage}[2]{{%
\expandafter\ifx\csname l@#1\endcsname\relax
\typeout{** WARNING: IEEEtran.bst: No hyphenation pattern has been}%
\typeout{** loaded for the language `#1'. Using the pattern for}%
\typeout{** the default language instead.}%
\else
\language=\csname l@#1\endcsname
\fi
#2}}
\providecommand{\BIBdecl}{\relax}
\BIBdecl

\bibitem{distr-alloc:it2012}
D.~Leong, A.~G. Dimakis, and T.~Ho, ``Distributed storage allocations,''
  \emph{Information Theory, IEEE Transactions on}, vol.~58, no.~7, pp. 4733
  --4752, july 2012.

\bibitem{heterog:isit12}
V.~Ntranos, G.~Caire, and A.~Dimakis, ``Allocations for heterogenous
  distributed storage,'' in \emph{Proceedings of ISIT'2012}, Boston, USA, July
  2012.

\end{thebibliography}
\end{document}